\documentclass[letter,longauth]{aa}
\usepackage{natbib}
\bibpunct{(}{)}{;}{a}{}{,} % to follow the A&A style
\usepackage{booktabs}
\usepackage{verbatim,multirow,multicol}
\usepackage{subfig}
\usepackage{hhline}
\usepackage{soul}
\setstcolor{red}
\usepackage{color}
\usepackage{tablefootnote}

\definecolor{Red}{rgb}{0.9,0,0}
\definecolor{Blue}{rgb}{0,0,0.9}
\definecolor{Green}{rgb}{0,0.5,0}
\definecolor{Black}{rgb}{0,0,0}

\newcommand{\Gaia}{\textit{Gaia}~}

\usepackage{txfonts}
\usepackage{graphicx}
\usepackage{multirow}
\usepackage{booktabs}
\usepackage[
colorlinks = true,
linkcolor = blue,
urlcolor  = black,
citecolor = blue]{hyperref}

\begin{document} 

\title{The stellar occultation by (319) Leona on 13 September 2023 in preparation for the occultation of Betelgeuse}
  
%\subtitle{subtitle}

   \author{J.~L.~Ortiz\inst{1} % email: ortiz@iaa.es,orcid: 0000-0002-8690-2413
    \and
    M.~Kretlow\inst{1,2}		% email: mike@kretlow.de,orcid: 0000-0001-8858-3420
    \and
    C.~Schnabel\inst{2,3}
    \and
    N.~Morales\inst{1}
    \and
    J.~Flores-Martín\inst{4}
    \and
    M.~Sánchez~González\inst{5}
    \and
    F. Casarramona\inst{3}
    \and
    A.~Selva\inst{2,3}
    \and
    C.~Perelló\inst{3} 
    \and
    A.~Román-Reche\inst{5} % same as A.~Román? 
    \and
    S.~Alonso\inst{6} 
    \and
    J.~L.~Rizos\inst{1} 
    \and
    R. Gonçalves\inst{2}
    \and
    A.~Castillo\inst{5}
    \and
    J.~M.~Madiedo\inst{1}
    \and
    P.~Martínez~Sánchez\inst{7}
    \and
    J.~M.~Fernández~Andújar\inst{8}
    \and
    J.~L.~Maestre\inst{9}
    \and
    E.~Smith\inst{10}
    \and
    M.~Gil\inst{10}
    \and
    V.~Pelenjow\inst{11}
    \and
    S.~Moral~Soriano\inst{12}
    \and 
    J.~Martí\inst{13}	% from universidad de Jaén
    \and 
    P.~L.~Luque-Escamilla\inst{13} 	% from universidad de Jaén
    \and
    R.~Casas\inst{3,14,15}	% ICE and Astrosabadell
    \and
    J.~Delgado Casal\inst{3}
    \and
    J.~Rovira\inst{3}
    \and
    F.~J.~Aceituno\inst{1}
    \and
    V.~Dekert\inst{16}
    \and
    V.~de~Ory~Guimerá\inst{17}
    \and	
    J. Serrano Estepa\inst{18}
    \and
    Y.~Kilic\inst{19}	% Tubitak email: yucelkilic1@gmail.com,orcid: 0000-0001-8641-0796
    \and
    R.~Leiva\inst{1}	%
    \and
    P.~Santos-Sanz\inst{1} % email: psantos@iaa.es,orcid: 0000-0002-1123-9830
    \and
    R.~Duffard\inst{1} % 
    \and
    E.~Fernández-Valenzuela\inst{20,1} % email: estela@ucf.edu,orcid: 0000-0003-2132-7769
    \and
    M. Vara-Lubiano\inst{1}
    \and
    A. Alvarez-Candal\inst{1,21}
    \and
    F.~L. Rommel\inst{22}
    }
    
    \institute{Instituto de Astrofísica de Andalucía – Consejo Superior de Investigaciones Científicas (IAA-CSIC), Glorieta de la Astronomía S/N, E-18008, Granada, Spain.~~\email{ortiz@iaa.es}
    \and International Occultation Timing Association - European Section (IOTA/ES), Am Brombeerhag 13, 30459 Hannover, Germany
    \and Agrupación astronómica de Sabadell, Prat de la Riba sn 08206 Sabadell, Spain
    \and Calar Alto Observatory, CAHA, Almeria, Spain.
    \and Sociedad Astronómica Granadina, Granada, Spain
    \and Universidad de Granada, Granada, Spain
\and Observatorio Casero La Alberca, Murcia, Spain.
\and Asociación de Astronomía de Sevilla
\and Albox Observatory, Almería, Spain.
\and Complejo Astronómico Los Coloraos, E-18890 Gorafe, (Granada) Spain
\and PixelSkies Castilléjar, Granada, Spain.
    \and Aula de Astronomía de Armilla, Granada, Spain
    \and Departamento de F\'{\i}sica, Escuela Polit\'ecnica Superior de Ja\'en, Universidad de Ja\'en, Campus Las Lagunillas s/n, A3-420, 23071 Ja\'en, Spain
    \and Institut d'Estudis Espacials de Catalunya, Gran Capità 2-4 Ed. Nexus, 08034, Barcelona, Spain
    \and Institute of Space Science, Carrer de Can Magrans, SN 08193 Cerdanyola de Vallès, Spain
    \and Observatorio de la Sagra, Granada, Spain.
    \and Observatorio de la Armada, San Fernando, Spain
    %\and Akdeniz University, Faculty of Sciences, Department of Space Sciences and Technologies, 07058 Antalya, Turkey
    \and Observatorio Fernán-Núñez, Córdoba, Spain.
    \and TÜB\.{I}TAK National Observatory, Akdeniz University Campus, 07058 Antalya, Turkey
    \and Florida Space Institute, UCF, 12354 Research Parkway, Partnership 1 building, Room 211, Orlado, USA
    \and Instituto de F\'isica Aplicada a las Ciencias y las Tecnolog\'ias, Universidad de Alicante, San Vicent del Raspeig, E03080, Alicante, Spain
    \and National Observatory/MCTI, R. General José Cristino 77, CEP20921-400, Rio de Janeiro, Brazil
%
\iffalse
    \and Institute for Astronomy and Astrophysics, Eberhard Karls University of Tübingen, Tübingen, Germany
    \and Florida Space Institute, UCF, 12354 Research Parkway, Partnership 1 building, Room 211, Orlado, USA
    \and Instituto de F\'isica Aplicada a las Ciencias y las Tecnolog\'ias, Universidad de Alicante, San Vicent del Raspeig, E03080, Alicante, Spain
\fi
    }

\date{Received September 20, 2023; accepted }

% \abstract{}{}{}{}{} 
% 5 {} token are mandatory
 
%\abstract
%% context heading (optional)
%% {} leave it empty if necessary  
%{}
%% aims heading (mandatory)
%{}
%% methods heading (mandatory)
%{}
%% results heading (mandatory)
%{}
%% conclusions heading (optional), leave it empty if necessary 
%{}
  
\abstract{On 12 December 2023, the star $\alpha$ Orionis (Betelgeuse) will be occulted by the asteroid (319) Leona. This represents an extraordinary and unique opportunity to analyze the diameter and brightness distribution of Betelgeuse's photosphere with extreme angular resolution by studying the light curve as the asteroid occults the star from different points on Earth and at different wavelengths. Here we present observations of another occultation by Leona on 13 September 2023 to determine its projected shape and size in preparation for the December 12th event. The occultation observation campaign was highly successful. The effective diameter in projected area derived from the positive detections at 17 sites turned out to be 66 km $\pm$ 2 km using an elliptical fit to the instantaneous limb. The body is highly elongated, with dimensions of 79.6 $\pm$ 2.2 km x 54.8 $\pm$ 1.3 km in its long and short axis, respectively, at the occultation time. Also, an accurate position coming from the occultation, to improve the orbit determination of Leona for December 12 is provided.}


\maketitle
%
%-------------------------------------------------------------------

\section{Introduction}
\begin{figure*}
\centering
% To include a figure from a file named example.*
% Allowable file formats are eps or ps if compiling using latex
% or pdf, png, jpg if compiling using pdflatex
%\includegraphics[width=\columnwidth]{experimentoFiguraChironV4.png}
\includegraphics[width=\textwidth]{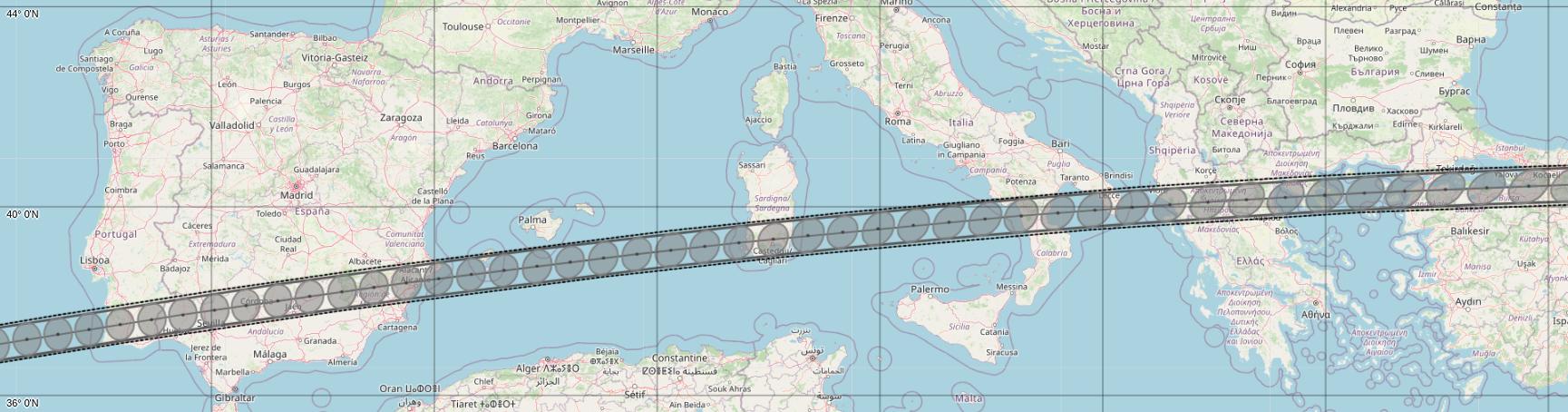}
\caption{Map of the shadow path over Europe predicted for the 12 December 2023 occultation of Betelgeuse by asteroid Leona. JPL Horizons \#69 ephemeris was used for Leona. The position of Betelgeuse was computed by using the epoch 2017.0 position from the  USNO-UBAD catalog \citep{Munn2022} and the proper motions were taken from \citet{Harper2017}. Map from OpenStreetMap.
%\footnote{ \url{https://astro.kretlow.de/cora/occultations/aa3579fb-2fa7-4a89-8452-bfc8a87e0704/}}
}
\label{Fig:Betelgeuse}
\end{figure*}

On 12 December 2023, the bright star $\alpha$ Orionis (Betelgeuse) will be occulted by the asteroid (319) Leona\footnote{This event was probably first found and announced by D.~Denissenko in 2004:  \url{https://web.archive.org/web/20121216061951/http://hea.iki.rssi.ru:80/~denis/special.html}}. The shadow path of this event will be favorable for a thin region crossing Southern Florida and Southern Europe. Even though this naked eye event will not be observable from the entire planet, highly populated areas of the Earth will be in the path, and the potential observability of the event is considerable. Figure~\ref{Fig:Betelgeuse} shows a map of the expected shadow path on Earth for the 12 December 2023 occultation of Betelgeuse using JPL$\#$69 orbit for Leona and a position for Betelgeuse that we have derived ourselves as explained below. Betelgeuse's position is somewhat problematic because being such a bright optical source there is no astrometry from Gaia. On the other hand, the Hipparcos astrometric solution for Betelgeuse is old and was questioned by \citet{Harper2017} who used radio observations to derive their own and more recent solution. However, there is the possibility that the central position of the radio emission from Betelgeuse be shifted with respect to the optical emission. Recently, ground-based astrometry of bright sources become available, with an accuracy of around 4 milliarcseconds (mas) by means of the USNO UBAD catalog \citep{Munn2022}, along with derivations of proper motions and parallax in a short time span compared to other works. For Figure ~\ref{Fig:Betelgeuse} we use the USNO UBAD position for Betelgeuse from 2017 propagated to December 2023 with the proper motions and parallax by \citet{Harper2017}. A clickable and expandable version of this map is available online\footnote{\url{https://astro.kretlow.de/cora/occultations/aa3579fb-2fa7-4a89-8452-bfc8a87e0704/}}. Also, we note that this prediction assumes a spherical Leona with a diameter of 61 km and uses an angular diameter of 41.9 $\pm$ 0.06 mas for Betelgeuse \citep{Dolan2016} which is actually the uniform disk Rossland mean diameter corrected for limb darkening provided in Table 3 of that paper. All this needs to be refined for an accurate prediction and interpretation of the occultation in December. 
Another interactive prediction map with a different orbit for Leona and other assumptions on Betelgeuse is also available online \footnote{\url{https://lesia.obspm.fr/lucky-star/occ.php?p=124370}}.

Asteroidal occultations of stars are frequent phenomena but the occultation of Betelgeuse by Leona will be extremely important and unique. It represents an extraordinary opportunity to analyze the diameter and brightness distribution of Betelgeuse's photosphere by studying the light curve as the asteroid progressively occults the star from different points on Earth and at different wavelengths. In the overwhelming majority of the stellar occultations the diameters of the stars are tiny compared to the angular size of the solar system body that passes in front of them so the occultations are not gradual, but in this case, the huge angular diameter of Betelgeuse will give rise to a different phenomenon of “partial eclipse” and “total eclipse” (provided that Leona's angular diameter is large enough compared with that of Betelgeuse). On the other hand, the interest in Betelgeuse has been extremely high in the last few years because of the large dimming that it experienced, which prompted all sorts of speculations and seems to be recently explained by a dusty veil \citep{Montarges2021}. Here we report results on Leona itself based on a stellar occultation on 13 September 2023 predicted and observed in preparation for the Betelgeuse event.

\section{Observations}
\begin{figure}
    \centering
    % To include a figure from a file named example.*
    % Allowable file formats are eps or ps if compiling using latex
    % or pdf, png, jpg if compiling using pdflatex
    \includegraphics[width=\columnwidth]{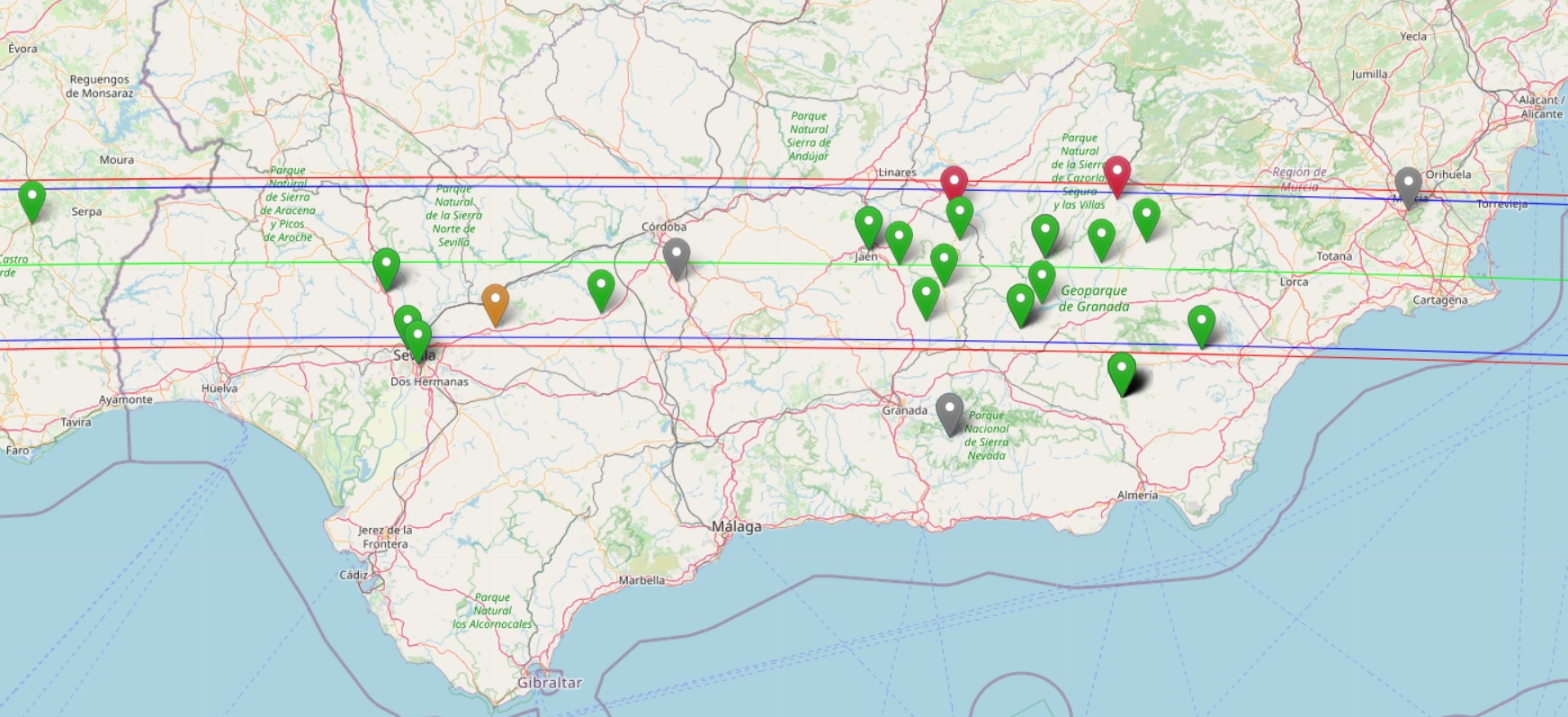} % TODO: replace with JPL69 CORA map
    \caption{Occultation shadow path computed for the 13 September 2023 stellar occultation by Leona prior to the event. The blue lines indicate the shadow path borders, the red lines indicate the uncertainty in the path coming from the orbit, the green line indicates the center of the shadow path, and the pins indicate the locations of the stations that participated in the campaign. The gray pins show stations that were clouded out or had weather problems, in green we show the sites that observed a positive occultation, in red the stations that recorded a negative and in brown the stations that had technical problems and could not observe. Map from OpenStreetMap.}
    \label{Fig:Leona13Sep}
\end{figure}

We predicted stellar occultations by Leona months in advance to the September event, first using ASTORB \citep{Mosko2022} orbital elements for Leona together with Gaia DR3 data for the stars, and once the most promising events in terms of observability were selected, we used the Jet Propulsion Laboratory (JPL) orbit available at the time to refine the predictions. Among the very best occultations in 2023, the  13 September 2023 event clearly stood up in terms of the brightness of the star involved and the logistic capabilities as well as the potential weather in the shadow path region. Therefore, we soon started organizing its observation. For the final deployment we followed a similar procedure to the one we use for stellar occultations by trans-Neptunian objects \citep{Ortiz2020b} and observations of Leona were obtained in an I filter to minimize Differential Chromatic Refraction, on 3 nights prior to the occultation with the 2\,m Liverpool Telescope on La Palma observatory, to try to refine Leona's shadow path prediction.
The final prediction map is shown in Figure \ref{Fig:Leona13Sep} together with the observing stations.
The details of the involved star are given in Table~\ref{Tab:StarDetails}.

In total, 23 observing stations participated in the campaign, 9 of which were specifically deployed for the event. All the participating stations are listed in Table~\ref{Tab:Observer} together with their instruments and setups used.

\begin{table}[!ht]
    \setlength{\tabcolsep}{2mm}  
    \centering
    \caption{
    Occulted star information. 
    }
    \begin{tabular}{l l}
    \toprule \toprule
    Epoch                               & 2023-09-13 03:42:00 UTC  \\
    Source ID                           & \Gaia DR3 3347400001862704896 \\
    Star position                       & $\alpha_\star$= 05$^h$42$^m$12$^s$.52922 $\pm$ 0.16 mas   \\
    \hspace{5mm}at epoch\tablefootmark{1}     & $\delta_\star$= 14$^\circ$00$'$04$''$.10756 $\pm$ 0.10 mas \\
    Magnitudes\tablefootmark{2}         &  G = 11.89; J = 10.34; H = 10.57; K = 10.23 \\
    Apparent diam.                   &  0.02 mas / 0.21 km \\
    \midrule
    \end{tabular}
    \tablefoot{
    \tablefoottext{1}{The star position was taken from the \Gaia Data Release 3 (GDR3) star catalog \citep{GaiaCollaboration2023}, and is propagated to the event epoch using the formalism of \citet{Butkevich2014} applied with the SORA package \citep{Gomes-Junior2022}. The duplicated source flag in GDR3 is 0, meaning that the star is not multiple or any companion would be very faint.
    \tablefoottext{2}{J, H and K magnitudes are taken from the 2MASS catalog.}
    }
    }
    \label{Tab:StarDetails}
\end{table}

\section {Data Reduction and Analysis}
\begin{figure}
    \centering
    % To include a figure from a file named example.*
    % Allowable file formats are eps or ps if compiling using latex
    % or pdf, png, jpg if compiling using pdflatex
    \includegraphics[width=\columnwidth]{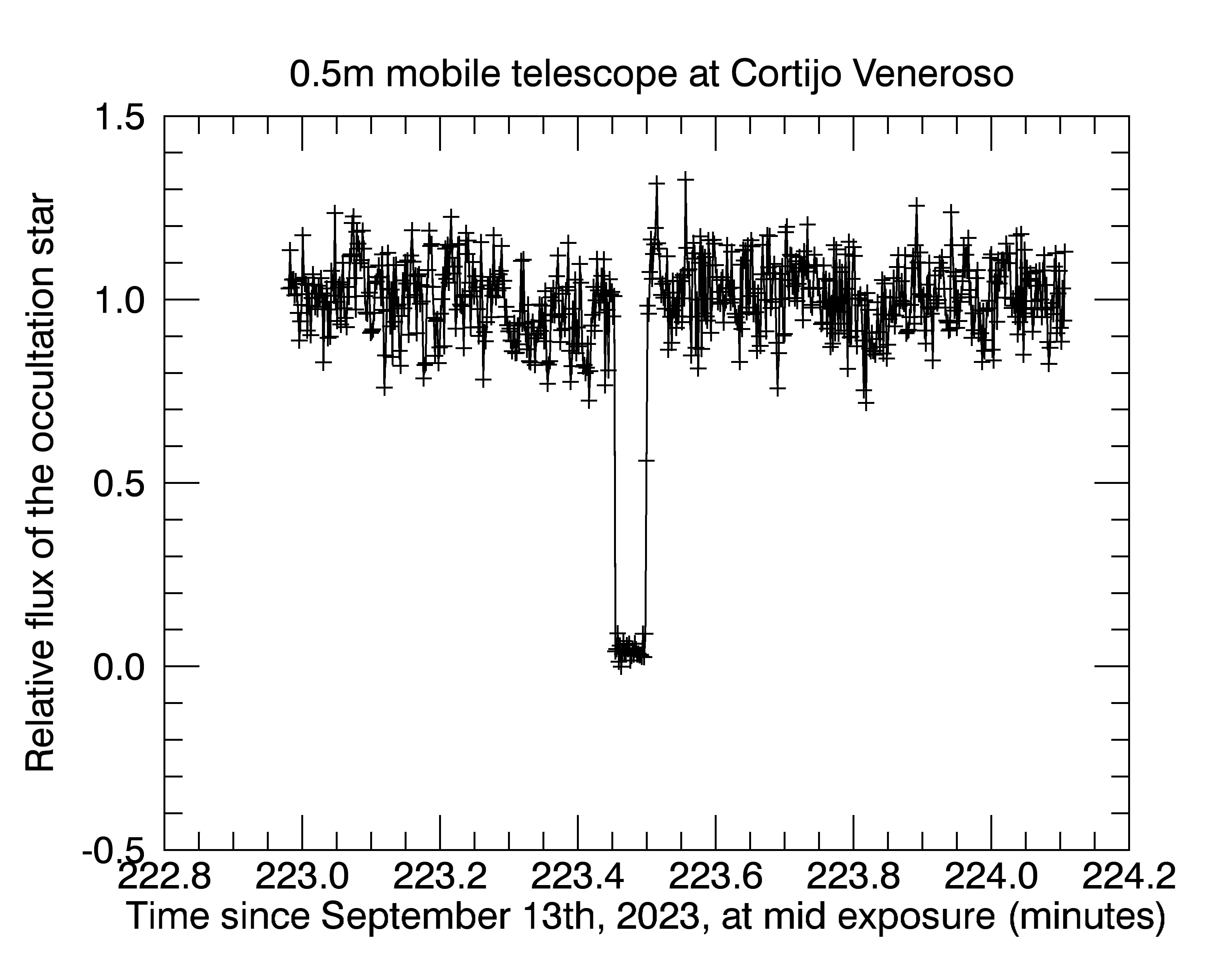} % 
    \caption{Light curve derived from the observations carried out at Cortijo Veneroso.}
    \label{Fig:Veneroso}
\end{figure}

% The data were managed through the Occultation portal website. \footnote{http://occultation.tug.tubitak.gov.tr}
The image sequences or the video files, compiled and managed through the Tubitak Occultation Portal website \citep{Kilic2022}, were dark-current subtracted and flatfielded whenever these calibration files were available. %Median bias frames and median sky flat frames were used.
Aperture photometry of the target star (blended with Leona) from the image sequences or video was carried out using different synthetic apertures to get the least dispersion possible in the relative photometry. Comparison stars of similar brightness were also measured to derive the relative photometry to compensate for small atmospheric fluctuations. The methods used were the same as those described in \cite{Ortiz2020b}. For a quick and uniform output of all the observations we used the occultation-oriented photometry software packages Tangra and Pymovie.
%(Ortiz et al. 2020). 
A sample light curve is shown in Figure \ref{Fig:Veneroso}. The rest of the light curves are provided online\footnote{\url{https://cloud.iaa.csic.es/index.php/s/QtesTtpJ7xpxE48}}.

The positive light curves show clear brightness drops from which time of ingress and time of egress at each site was derived. This was done through square well fits to the occultation profiles as usually done in occultation work (e.g. \cite{Ortiz2020}, \cite{Morgado2023}). For a fast and uniform derivation of the ingress and egress times we used the Asteroidal Occultation Time Analyser (AOTA) and Python Occultation Timing Extractor (PyOTE) software packages. The ingress and egress times for all the sites are listed in Table ~\ref{Tab:Timings}. Due to the urgency to report results, conservative errors were used for the ingress and egress times. Three of the stations had gaps (deadtime) in the image acquisition sequence.  In these cases we used half the deadtime as the uncertainty, to be on a safe side. One of the stations with gaps in the acquisition was at the Calar Alto observatory, but given that we used seven 0.4m telescopes there, two of the telescopes captured the ingress during the integration time so the ingress time could be determined accurately.

One of the stations recorded the occultation from a trailed image using the drift-scan technique. In this case the ingress and egress times were derived with scanalyzer software\footnote{\url{http://www.asteroidoccultation.com/observations/DriftScan/Scanalyzer.zip}}, which is specific for this type of observations.

Besides, for those instruments using NTP as the UT synchronization tool for the acquisition computers we included an additional 0.1s uncertainty which we added quadratically for the analysis in the next section. Three stations did not have their computers properly synchronized through NTP due to problems in the configuration. The resulting chords from these stations deviated somewhat from nearby chords so we applied small offsets to align with the nearby chords. These stations were not used for the elliptical fit shown in the next section.

The ingress and egress times will be revised and their uncertainties decreased in the next two months prior to the December Betelgeuse occultation and the results will be updated, but given the large number of chords on the body and the short integrations used, the main results are not expected to change significantly.

%The observations at the Inmaculada del Molino station showed problems in the timing and other issues so it has not been analyzed here. It is included in Table  ~\ref{Tab:Timings} with entries ``TBD'' meaning that we are working on the data. We intend to find the origin of the problems and fix them, but so far only 16 chords have been included in the plots and in the analysis.

%From the light curves we extracted the times at which the sharp drops

\section{Interpretation}
%INCLUDE THE FINAL CHORDS IN THE PLANE OF THE SKY, AN ELLIPTICAL FIT AND THE DERIVED ASTROMETRIC OFFSET. THIS SHOULD BE ENOUGH FOR A QUICK PAPER.

%WITH ASTROMETRY CALCULATE NEW OWN ORBIT AND PREDICTION USING HARPER POSITION FOR BETELGEUSE:

\begin{figure}
    \centering
    % To include a figure from a file named example.*
    % Allowable file formats are eps or ps if compiling using latex
    % or pdf, png, jpg if compiling using pdflatex
    \includegraphics[width=\columnwidth]{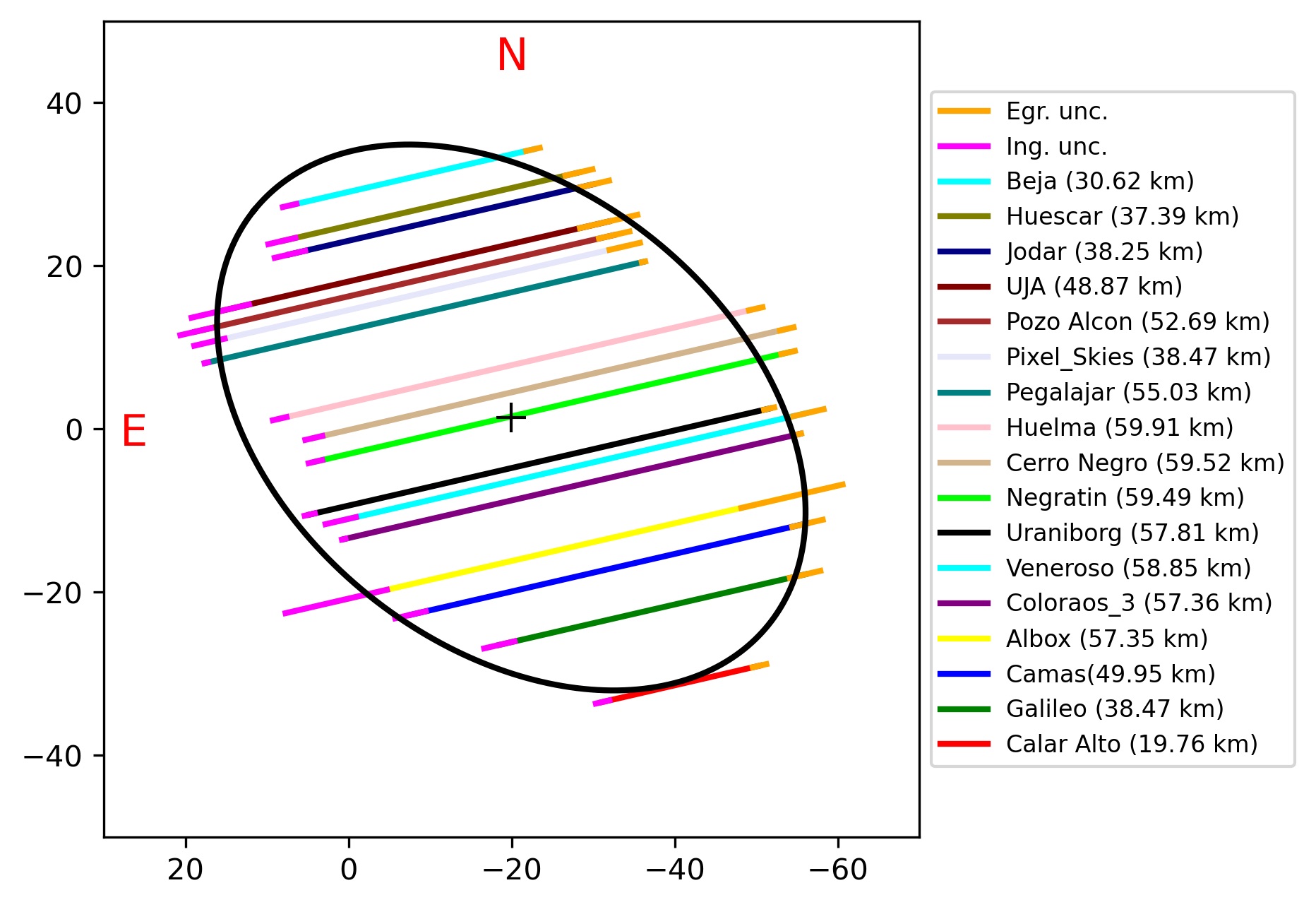} % 
    \caption{Chords of the stellar occultation and an elliptical fit to the extremities. The ingress uncertainties are shown with pink color and the egress uncertainties in clear brown. The values in parenthesis are the chords lengths. The scale is in km. The chords were built using the JPL$\#$69 ephemeris.}
    \label{Fig:Chords}
\end{figure}

To analyze the projected shape of Leona we first used the set of measurements that were obtained with GPS time insertion to make sure we do not have effects due to timing issues, and also used the negative occultation north of the body together with the grazing occultation at Calar Alto observatory, near the south part of the body. From these, we obtained the least squares elliptical fit shown in Figure \ref{Fig:Chords}. The dimensions are 78.2 $\pm$ 2.5 km x 54.0 $\pm$ 1.3 km with a position angle of 55 $\pm$ 5 degrees for the major axis of the ellipse. We note that the rms of the fit is 2.3 km, meaning that an ellipsoidal shape is an overall rough approximation for Leona´s shape as the residuals are slightly above those expected from the accuracy in the ingress and egress timing of the GPS-based chords. We note that even though the Huelma chord was obtained with a GPS-based camera, the GPS did not get a lock so NTP sync was finally used and this station was not included in the initial fit.

Then, we incorporated the rest of the chords and performed another fit, which did not change the results of the fitted ellipse within the error bars. The dimensions are 79.6 $\pm$ 2.2 km and 54.8 $\pm$ 1.3 km with a position angle of 50.6 $\pm$ 3.5 degrees for the long axis of the ellipse. The rms of the fit was 2.6 km, also very similar to the fit of the GPS-based chords. From visual inspection there appears to be deformations of the body with respect to an ellipsoidal shape. From the fit, the center of the body was off with respect to the JPL$\#$69 ephemeris by -19.9 $\pm$ 0.6 km and +1.5 $\pm$ 0.8km in RA*cos(Dec) and Declination respectively, or -10.2 $\pm$ 0.3 mas and 0.8 $\pm$ 0.4 mas. 
From the fit, the geocentric position of Leona at time 2023-09-13 03:46:29.440 was RA = 05h42m12.54493s $\pm$ 0.4 mas; Dec = 14$^\circ$00' 05.1461" $\pm$ 0.5 mas.

The effective projected-area diameter from the elliptical fit is 66 $\pm$ 2 km. This is considerably larger than the effective thermal diameter of 49.943 $\pm$ 0.477 km reported in the WISE/NEOWISE catalog \citep{Masiero2014} while very close to the 65.90 $\pm$ 0.92 km from AKARI measurements reported in a previous paper by \citet{Usui2011}. Our determination is also consistent with the much coarser thermal diameter estimate of 89.00 $\pm$ 27.92 from the NEOWISE reactivation mission as reported in \citet{Nugent2016}.

In its longest dimension Leona is considerably larger than the nominal 61 km value that was adopted here to make the prediction maps. This axis, projected in the direction perpendicular to the chord will be 60 km on December 13, corresponding to 46 mas. This means that the area of annularity of the December occultation would be near the nominal prediction in Figure 1. On the other hand, in the small dimension the 54\,km translate into 41 mas on December 13th, just very near the angular size for Betelgeuse according to \citep{Dolan2016}, and near the limit to produce totality.

As a caveat to these statements we must note that the rotational phase at which the occultation took place is not well known. In an online report \cite{Behrend2023} shows a $2.14533\pm0.00026$ day rotation period with a light curve amplitude of around 0.5\,mag\footnote{see \url{https://obswww.unige.ch/~behrend/r000319-01a.png}}. However, \citet{Pilcher2017} provided two long period solutions ($430 \pm 2$ and $1084 \pm 10$ hours), one of which could be indicating tumbling and reported a $\sim$ 0.7\,mag amplitude. This appears difficult to reconcile with the other results.

Given the 0.5\,mag to 0.7\,mag amplitude of the rotational light curve it is possible that the September occultation happened near a maximum of the rotation and therefore the dimensions at the time of the December occultation might be different.

If the period of 2.14533~days is maintained with no alteration from a longer period there would be a 0.9 rotational phase difference between the 12 December 2023 event and the September occultation reported here. Therefore, an important effort to determine rotational light curves prior to the December event is necessary.

Using the zero-phase epoch and the period in \cite{Behrend2023} the phase at the time of the September occultation was 0.64 which means that the body was near its mean projected size. As already stated, we obtained images of Leona with the 2\,m Liverpool telescope on three nights, but the time coverage was insufficient to determine a rotational light curve and unfortunately our monitoring was interrupted due to a fault in the telescope enclosure so we cannot derive a rotational light curve right now.

Also, regarding the angular diameter of Betelgeuse, we note that this depends on wavelength and on numerous aspects as discussed in \cite{Dolan2016}. Hence, observations of the Betelgeuse occultation are encouraged even for observers somewhat outside the nominal shadow path.

\section{Conclusions}
We successfully observed a stellar occultation by Leona on 13 September 2023. A large number of chords (17) on the body were obtained and a near miss to the north was also recorded, obtaining an excellent body coverage. This allowed us to derive accurate dimensions for Leona and an approximate silhouette. From an elliptical fit to the instantaneous limb the dimensions are 79.6 $\pm$ 2.2 km and 54.8 $\pm$ 1.3 km. Small deviations from an ellipsoidal shape are observable in the limb with rms of 2.6 km. The effective diameter in area sense is 66 km $\pm$ 2 km, somewhat larger than the previous expectations, but the rotational phase at the time of the occultation must be taken into account to make conclusions on the apparent limb expected for the 12 December 2023 event. Given that the amplitude of the rotational light curve is around 0.5\,mag, this means that our derivation can be up to 50\% off if Leona was near a maximum at 3:43 UT on 13 September 2023. Hence, rotational light curves need to be obtained. Also, an accurate position for the center of the body has been obtained with an accuracy below the 1 milliarcsecond level from which the orbit of Leona can be improved in preparation for the December Betelgeuse event.

\begin{acknowledgements}
Part of this work was supported by the Spanish projects PID2020-112789GB-I00 from AEI and Proyecto de Excelencia de la Junta de Andalucía PY20-01309. Financial support from the grant CEX2021-001131-S funded by MCIN/AEI/ 10.13039/501100011033 is also acknowledged.
JM and PLLE are supported by grant PID2019-105510GB-C32 / AEI / 10.13039/501100011033 from the State Agency for Research of the Spanish Ministry of Science and Innovation. They also acknowledge support by Consejer\'{\i}a de Econom\'{\i}a, Innovaci\'on, Ciencia y Empleo of Junta de Andaluc\'{\i}a as research group FQM- 322, as well as FEDER funds. P.S-S. acknowledges financial support from the Spanish project PID2022-139555NB-I00. We thank M. Pugnaire for observing support at Los Coloraos observatory. A.R. and S.A. acknowledge the financial support through the Europlanet Society Public Engagement Funding Scheme 2023, sponsored by the 'Dill Faulkes Educational Trust'. This article is partly based on observations made in the Observatorios de Canarias of IAC with the Liverpool Telescope operated on the island of La Palma by the Liverpool JMU in the Observatorio del Roque de los Muchachos. 
\end{acknowledgements}

% WARNING
%-------------------------------------------------------------------
% Please note that we have included the references to the file aa.dem in
% order to compile it, but we ask you to:
%
% - use BibTeX with the regular commands:
%   \bibliographystyle{aa} % style aa.bst
%   \bibliography{Yourfile} % your references Yourfile.bib
%
% - join the .bib files when you upload your source files
%-------------------------------------------------------------------
% 
\bibliographystyle{aa}
\bibliography{references.bib}

% Appendix A
\onecolumn
\appendix
\section{Observational data}

% This table is generated by CORA, so do not change/modify here, make changes in CORA.
% This table format does not contain the comments field. But the ASCII/PSV format does.

\begin{footnotesize}
\begin{longtable}{rllllllllll}
\caption{Observation details. Given are the sitename (or nearest city), the observer name(s), the country code CC (of the observing site), the geodetic coordinates of the site, the used telescope / optics (Mxxx means mirror telescope with xxx mm aperture), the detector / camera model, whether a filter was used and which one, the observation / recording method (IMA: digital imaging recording, VID: video camera, i.e. analog imaging recording, DRS: drift scan method), the time source used for the timing of the recording, and the exposure time; Occ describes the result / status of the observation (O+: positive detection recorded, O-: negative / no detection, ND: no data obtained (e.g. technical failure, bad weather, etc.)).}
\label{Tab:Observer}
\\\toprule
 \# & Sitename & CC & Latitude (dms) & Optics & Method & Occ \\
   & Observer(s) &  &  Longitude (dms) & Detector & Timesrc & Note \\
   &             &  &  Elevation       & Filter   & ExpTime &      \\\midrule
0  & Fernán Núñez                   & ES & N 37 39 56.95 & M255   & IMA & ND  \\& & & W 04 43 10.96 & CANON 450 & NTP &\\& \multicolumn{1}{l}{\parbox[t]{5cm}{\raggedright\em J. Serrano Estepa}} & & 344\,m  &  &  \\\midrule		
1  & Sierra Nevada Observatory      & ES & N 37 03 50.88 & M900   & IMA & ND  \\& & & W 03 23 04.92 & QHY600    & NTP &\\& \multicolumn{1}{l}{\parbox[t]{5cm}{\raggedright\em F.J. Aceituno, J.L. Ortiz, M. Kretlow}} & & 2931\,m &  &  \\\midrule
2  & Carmona                        & ES & N 37 29 18.4  & M203   & IMA & ND  \\& & & W 05 36 06    & QHY174GPS & GPS &\\& \multicolumn{1}{l}{\parbox[t]{5cm}{\raggedright\em V. de Ory Guimerá}} & &   67\,m &  &  \\\midrule
3  & Casero - La Alberca            & ES & N 37 56 29.31 & M200   & IMA & ND  \\& & & W 01 08 36.11 & Canon R5C (Super16mm) & NTP &\\& \multicolumn{1}{l}{\parbox[t]{5cm}{\raggedright\em P. Martínez Sánchez}} & &   60\,m & None & 0.1s \\\midrule
4  & Calar Alto (CAHA): MARCOT M5   & ES & N 37 13 24.71 & M406   & IMA & O+  \\& & & W 02 32 44.88 & ZWO ASI1600mm pro & GPS &\\& \multicolumn{1}{l}{\parbox[t]{5cm}{\raggedright\em J. Flores}} & & 2173\,m & Luminance & 0.1s \\\midrule
5  & Galileo                        & ES & N 37 20 46.42 & M280   & IMA & O+  \\& & & W 05 58 48.72 & QHY174M-GPS & GPS &\\& \multicolumn{1}{l}{\parbox[t]{5cm}{\raggedright\em J.M. Madiedo }} & &   11\,m & None & 0.1s \\\midrule
6  & Sevilla (Camas)                & ES & N 37 24 11.3  & M280   & IMA & O+  \\& & & W 06 01 55.7  & QHY174GPS & GPS &\\& \multicolumn{1}{l}{\parbox[t]{5cm}{\raggedright\em J. Delgado}} & &  120\,m & None & 0.15 \\\midrule
7  & Albox                          & ES & N 37 24 20.03 & M280   & IMA & O+  \\& & & W 02 09 06.48 & Atik314L+ & NTP &\\& \multicolumn{1}{l}{\parbox[t]{5cm}{\raggedright\em J.L. Maestre}} & &  491\,m & Clear & 2.00s \\\midrule
8  & Los Coloraos, Dome C2          & ES & N 37 29 10.98 & M280   & IMA & O+  \\& & & W 03 02 17.27 & ZWO ASI174 MM & NTP &\\& \multicolumn{1}{l}{\parbox[t]{5cm}{\raggedright\em M. Gil, E. Smith, F. Casarramona}} & &  990\,m & None & 0.08043s \\\midrule
9 & Los Coloraos, Dome C3          & ES & N 37 29 11.06 & M235   & VID & O+  \\& & & W 03 02 17.21 & Watec 910HX/RC & GPS &\\& \multicolumn{1}{l}{\parbox[t]{5cm}{\raggedright\em F. Casarramona, E. Smith, M. Gil}} & &  990\,m & None & 0.02s \\\midrule
10 & Los Coloraos, Dome C4          & ES & N 37 29 11.18 & M356   & IMA & O+  \\& & & W 03 02 17.19 & ZWO ASI6200 MM Pro & NTP &\\& \multicolumn{1}{l}{\parbox[t]{5cm}{\raggedright\em E. Smith, F. Casarramona, M. Gil}} & &  990\,m & None & 0.0352910s \\\midrule
11 & Cortijo Veneroso               & ES & N 37 30 47.77 & M500   & IMA & O+  \\& & & W 03 29 56.4  & QHY174M-GPS & GPS &\\& \multicolumn{1}{l}{\parbox[t]{5cm}{\raggedright\em N. Morales, J.L. Rizos, M. Kretlow}} & & 1137\,m & None & 0.1s \\\midrule
12 & Observatorio Uraniborg         & ES & N 37 32 33    & M280   & DRS & O+  \\& & & W 05 05 05.64 & Atik 414EX Mono & NTP &\\& \multicolumn{1}{l}{\parbox[t]{5cm}{\raggedright\em R.G. Farfan}} & &  110\,m & CV &  \\\midrule
13 & Mirador del Negratín           & ES & N 37 34 51.18 & M280   & VID & O+  \\& & & W 02 56 06.84 & Watec 910HX/RC & GPS &\\& \multicolumn{1}{l}{\parbox[t]{5cm}{\raggedright\em A. Selva }} & &  833\,m & None & 0.08s \\\midrule
14  & Cerro Negro, Sevilla          & ES & N 37 37 34.39 & M200   & IMA & O+  \\& & & W 06 07 59.80 & ZWO ASI1600 & GPS &\\& \multicolumn{1}{l}{\parbox[t]{5cm}{\raggedright\em J.M. Fernández Andújar}} & &   295\,m & None & 0.1s \\\midrule
15 & Huelma                         & ES & N 37 38 40.38 & M280   & IMA & O+  \\& & & W 03 24 46.92 & QHY174M-GPS & GPS &\\& \multicolumn{1}{l}{\parbox[t]{5cm}{\raggedright\em J. Flores}} & &  980\,m & None & 0.1s \\\midrule
16 & Pegalajar                      & ES & N 37 43 48.73 & M635   & IMA & O+  \\& & & W 03 37 42.70 & ASI 174 MM Mono & NTP &\\& \multicolumn{1}{l}{\parbox[t]{5cm}{\raggedright\em A. Román-Reche, S. Alonso}} & &  769\,m & None & 0.1s \\\midrule
17 & PixelSkies, Castilléjar         & ES & N 37 44 24.07 & M279   & IMA & O+  \\& & & W 02 38 38.22 & ASI2400MC Pro & GPS &\\& \multicolumn{1}{l}{\parbox[t]{5cm}{\raggedright\em V. Pelenjow}} & &  850\,m & UV/IR & 0.35s \\\midrule
18 & Pozo Alcón                     & ES & N 37 45 32.19 & M203   & VID & O+  \\& & & W 02 54 54.29 & Watec 910HX/RC & GPS &\\& \multicolumn{1}{l}{\parbox[t]{5cm}{\raggedright\em C. Perelló, C. Schnabel}} & & 1050\,m & None & 0.08s \\\midrule
19 & UJA Observatory                & ES & N 37 47 14.46 & M406   & IMA & O+  \\& & & W 03 46 39.68 & Atik 314L & NTP &\\& \multicolumn{1}{l}{\parbox[t]{5cm}{\raggedright\em J. Martí, P.L. Luque-Escamilla}} & &  562\,m & None & 0.1s \\\midrule
20 & Jodar                          & ES & N 37 49 43.33 & M260   & IMA & O+  \\& & & W 03 20 02.49 & Player One Apollo-M Max & NTP &\\& \multicolumn{1}{l}{\parbox[t]{5cm}{\raggedright\em M. Sánchez González, S. Moral Soriano}} & &  630\,m & Clear & 0.3s \\\midrule
21 & Casas de don Juan, Huescar     & ES & N 37 49 5.23  & M280   & VID & O+  \\& & & W 02 25 18.89 & Watec 910HX & GPS &\\& \multicolumn{1}{l}{\parbox[t]{5cm}{\raggedright\em J. Rovira}} & &  990\,m & Clear & 0.16s \\\midrule
22 & Beja                           & PT & N 37 53 17.98 & M256   & VID & O+  \\& & & W 07 51 59.00 & Watec910HX-RC & GPS &\\& \multicolumn{1}{l}{\parbox[t]{5cm}{\raggedright\em R. Gonçalves}} & &  155\,m & Clear & 0.08s \\\midrule
23 & Úbeda                          & ES & N 37 56 37.15 & M280   & IMA & O-  \\& & & W 03 21 41.61 & ASI1600MM Pro Mono & NTP &\\& \multicolumn{1}{l}{\parbox[t]{5cm}{\raggedright\em A. Castillo}} & &  368\,m & None & 0.2s \\\midrule
24 & Observatorio de La Sagra       & ES & N 37 58 57.39 & M356   & IMA & O-  \\& & & W 02 33 57.6  & QHY174-GPS & GPS &\\& \multicolumn{1}{l}{\parbox[t]{5cm}{\raggedright\em  V. Dekert, N. Morales, J. L. Ortiz}} & & 1530\,m & None & 0.1s \\\midrule
\end{longtable}
\end{footnotesize}

%\clearpage

\begin{small}
\begin{longtable}{rlllllrl}
\caption{Measured occultation timings}
\label{Tab:Timings}
\\\toprule
\# & Sitename             & Ingress (UT)& I.err   & Egress (UT) & E.err   & Duration   & Note \\\midrule
4  & Calar Alto (CAHA): MARCOT M5   & 03:43:31.17 & 0.04\,s & 03:43:32.1  & 0.3 \,s &    0.93\,s & \\
5  & Galileo, Sevilla               & 03:43:20.02 & 0.09\,s & 03:43:21.83 & 0.09\,s &    1.81\,s & \\
6  & Sevilla (Camas)                & 03:43:19.46 & 0.09\,s & 03:43:21.81 & 0.09\,s &    2.35\,s & \\
7  & Albox                          & 03:43:31.3  & 0.3 \,s & 03:43:34.0  & 0.3 \,s &    2.70\,s & \\
8  & Los Coloraos, Dome C2          & 03:43:28.75 & 0.01\,s & 03:43:31.45 & 0.01\,s &    2.70\,s & \\
9  & Los Coloraos, Dome C3          & 03:43:28.63 & 0.01\,s & 03:43:31.33 & 0.01\,s &    2.70\,s & \\
10 & Los Coloraos, Dome C4          & 03:43:28.71 & 0.00\,s & 03:43:31.40 & 0.00\,s &    2.69\,s & \\
11 & Cortijo Veneroso               & 03:43:27.19 & 0.09\,s & 03:43:29.96 & 0.09\,s &    2.77\,s & \\
12 & Observatorio Uraniborg         & 03:43:22.10 & 0.03\,s & 03:43:24.82 & 0.03\,s &    2.72\,s & \\
13 & Mirador del Negratín           & 03:43:29.00 & 0.04\,s & 03:43:31.80 & 0.04\,s &    2.80\,s & \\
14 & Cerro Negro, Sevilla           & 03:43:19.06 & 0.05\,s & 03:43:21.86 & 0.04\,s &    2.80\,s & \\
15 & Huelma                         & 03:43:27.39 & 0.04\,s & 03:43:30.21 & 0.04\,s &    2.82\,s & \\
16 & Pegalajar                      & 03:43:26.46 & 0.01\,s & 03:43:29.05 & 0.01\,s &    2.59\,s & \\
17 & PixelSkies, Castilléjar        & 03:43:29.65 & 0.16\,s & 03:43:32.11 & 0.17\,s &    2.46\,s & \\
18 & Pozo Alcón                     & 03:43:28.73 & 0.09\,s & 03:43:31.21 & 0.09\,s &    2.48\,s & \\
19 & UJA Observatory                & 03:43:26.20 & 0.17\,s & 03:43:28.50 & 0.17\,s &    2.30\,s & \\
20 & Jodar                          & 03:43:28.11 & 0.09\,s & 03:43:29.91 & 0.09\,s &    1.80\,s & \\
21 & Casas de don Juan, Huescar     & 03:43:30.96 & 0.08\,s & 03:43:32.72 & 0.08\,s &    1.76\,s & \\
22 & Beja                           & 03:43:14.35 & 0.04\,s & 03:43:15.79 & 0.04\,s &    1.44\,s & \\
\\\midrule
\end{longtable}
\end{small}

\end{document}